\documentclass{aastex62}

\newcommand{\mj}{M_\mathrm{J}}
\newcommand{\mearth}{M_{\earth}}
\newcommand{\msun}{M_{\odot}}
\newcommand\rj{R_\mathrm{J} }
\newcommand\re{R_{\earth} }
\newcommand{\rh}{R_\mathrm{H}}

\newcommand\mpl{M_\mathrm{p} }
\newcommand\vescp{v_\mathrm{esc,p}}
\newcommand\mdisk{M_\mathrm{disk}}
\newcommand\mmed{1.17}
\usepackage{amsmath}

\received{June 7, 2019}
\revised{September 18, 2019}
\accepted{October 2, 2019}

\submitjournal{ApJL}

\shortauthors{Frelikh et al.}

\begin{document}

\title{Signatures of a planet-planet impacts phase in exoplanetary systems hosting giant planets}

\correspondingauthor{Renata Frelikh}
\email{rfrelikh@ucsc.edu}

\author[0000-0003-3290-818X]{Renata Frelikh}
\affil{Department of Astronomy and Astrophysics \\
UCO/Lick Observatory, University of California at Santa Cruz \\
Santa Cruz, CA 95064, USA}

\author{Hyerin Jang}
\affil{Department of Astronomy and Astrophysics \\
UCO/Lick Observatory, University of California at Santa Cruz \\
Santa Cruz, CA 95064, USA}

\author[0000-0001-5061-0462]{Ruth A. Murray-Clay}
\affil{Department of Astronomy and Astrophysics \\
UCO/Lick Observatory, University of California at Santa Cruz \\
Santa Cruz, CA 95064, USA}

\author[0000-0003-0412-9314]{Cristobal Petrovich}
\affiliation{Canadian Institute for Theoretical Astrophysics \\
University of Toronto \\
60 St. George Street, Toronto, ON M5S 3H8, Canada}

\begin{abstract}

Exoplanetary systems host giant planets on substantially non-circular, close-in orbits. We propose that these eccentricities arise in a phase of giant impacts, analogous to the final stage of Solar System assembly that formed Earth's Moon. In this scenario, the planets scatter each other and collide, with corresponding mass growth as they merge. We numerically integrate an ensemble of systems with varying total planet mass, allowing for collisional growth, to show that (1) the high-eccentricity giants observed today may have formed preferentially in systems of higher initial total planet mass, and (2) the upper bound on the observed giant planet eccentricity distribution is consistent with planet-planet scattering. We predict that mergers will produce a population of high-mass giant planets between 1 and 8 au from their stars.

\end{abstract}


\section{Introduction} \label{intro}
Observations of exoplanetary systems have found many gas giants with orbital distances less than 5 au from their host stars. The orbits of these planets are often eccentric, deviating from the roughly circular orbits in our Solar System. Several mechanisms have been proposed to account for these eccentricities, including: secular chaos \citep{wu}, the Kozai-Lidov mechanism \citep[e.g.][]{dawson2014,kozai,lidov,naoz,takeda}, secular oscillations due to an outer planetary companion \citep{petr2016}, resonant interactions with a gas disk \citep{chiang}, and planet-planet scattering \citep[e.g.][]{chatterjee,juric,petr2014,rasio,ford2001,ford2008}. However, these studies have not yet explained an important clue to these systems' dynamical histories: most observed planets with eccentricities $e>0.5$ are more massive than Jupiter, while lower-mass planets are confined to lower eccentricities \citep{winn}. This observation is potentially surprising, since lower-mass planets are excited to higher eccentricities in typical dynamical simulations \citep[e.g.][Figures 14 and 20]{chatterjee}.

We explore the possibility that some stars initially host multiple hydrogen-rich planets in their inner systems, which go through a giant-impacts phase, analogous to the final stage of inner Solar System assembly that resulted in Earth's moon-forming impact. Collisions cause these planets to grow in mass. Though in a given system, the lower-mass planets are more likely to have higher eccentricities, when viewed as a population, the trend reverses. Stars hosting more total mass in planets produce more high-mass giants, which in turn are able to excite each other to high orbital eccentricities. We use the observed mass distribution for planets at semimajor axes $a < 5$ au to construct a distribution of initial total planet masses in that region. We demonstrate, using stellar metallicity as an observational proxy for initial mass in planets, that our ensemble produces giant-impact phase outcomes consistent with observations.

\section{Observational Sample} \label{sec:sample}
We compare our simulations to a sample of observed exoplanets obtained from the Exoplanet Orbit Database on April 3, 2019, hosted on exoplanets.org \citep{wright}. We include 311 planets from 243 stellar systems, discovered via the radial-velocity method, orbiting FGK stars (0.5 to 1.4 $M_\sun$). Following \citet{dawson2013}, we remove 70 evolved stars with log g $<4$. We exclude 11 planets for which no data for log g is reported. We further exclude 27 planets with no reported eccentricity. We note that all of the masses in the observational data set are reported as $M\mathrm{sin}i$, which henceforth we refer to as the mass. The planets in this sample are subject to several biases, including radial-velocity selection biases toward higher-mass, close-in planets \citep{gaudi2005a,gaudi2005b}. It is not appropriate to make a statistical comparison between the simulated and the observed data sets, as there is a risk that the observational sample used is non-uniform. We therefore focus on the qualitative features of the distribution.

\section{The Highest-Eccentricity Planets are Found Around Metal-rich Stars}
\begin{figure*}[ht!] 
\plotone{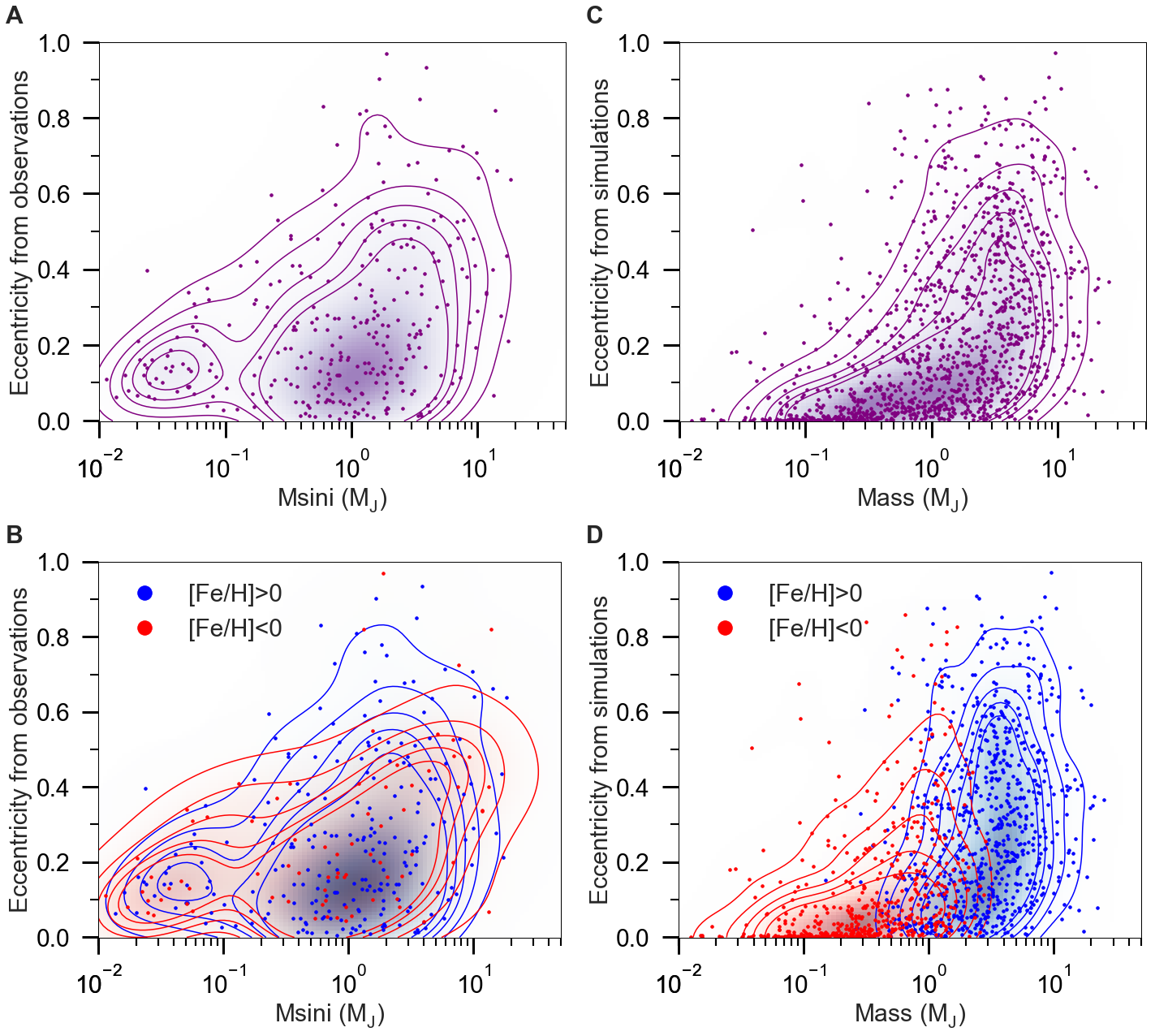}
\caption{Distribution of eccentricities as a function of planet mass for (A) \& (B) observations, and (C) \& (D) simulations. The eccentricity of observed exoplanets is correlated with planet mass (A), which is recovered in our simulations (C). In (B), planets orbiting metal-rich ([Fe/H]$>$0, blue, 235 points) stars exhibit a range of eccentricities. Planets orbiting metal-poor stars ([Fe/H]$<$0, red, 76 points) are confined to $e < 0.6$. The highest-eccentricity planets ($e > 0.6$) are giant planets with $\mpl > 0.5$ Jupiter masses ($\mj$). This is matched in (D), in which we use total initial mass in planets as a proxy for metallicity (see Section \ref{disk}). On all panels, the upper contour denotes an enclosed probability of 90\%, and each successive contour below is a 10\% decrement to the one above.} \label{f1}
\end{figure*}

Figure \ref{f1}A displays the distribution of eccentricity as a function of planetary mass for the observational sample. The highest-eccentricity planets tend to be more massive, contradicting our intuition that the lower masses are more readily excited. Figure \ref{f1}B provides a hint to the solution. After dividing the data into planets orbiting metal-rich stars ([Fe/H]$>$0, blue) and metal-poor stars ([Fe/H]$<$0, red), we find that planets with the highest observed eccentricities orbit stars that are preferentially metal-rich\footnote{The Solar [Fe/H]$=$0 cutoff is employed because the occurrence rate of giant planets rises steeply for stellar metallicities above [Fe/H]$=$0 \citep{santos,fischer}. 76 planets are observed in systems with [Fe/H]$<$0, and 235 - in systems with [Fe/H]$>$0. The median metallicity in our observational sample is [Fe/H]$=$0.14.}. Therefore, if metal-rich stars tend to produce a greater quantity of high-mass planets, these giants would scatter their planetary companions to higher eccentricities. This is a key point of our Letter.

\begin{figure*}[ht!] 
\plotone{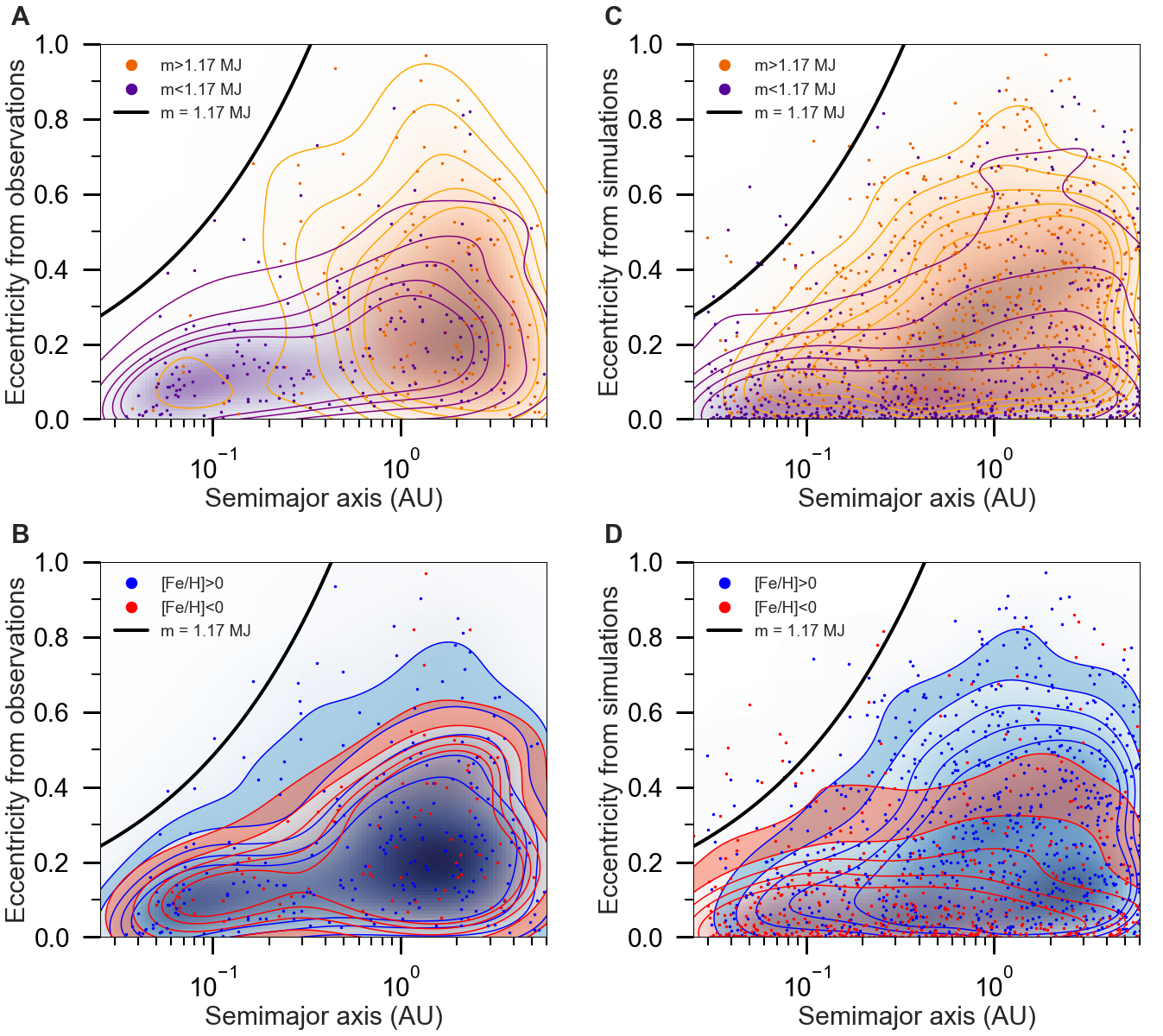}
\caption{Eccentricity vs. semimajor axis. (A) Observational eccentricity distribution. The high-mass giant planets ($m > \mmed$ $\mj$, orange) exhibit eccentricities $0 < e \lesssim 1$, and semimajor axes generally beyond 0.3 au, while low-mass giants (purple) span 0.03-6 au in $a$. This is matched by the simulations in (C), though the mass segregation in $a$ is weaker. The eccentricity envelope (black solid line) denotes the planet-planet scattering limit to the eccentricity (Equation \ref{envelope}). In (B) and (D), we divide the planets into those belonging to systems orbiting metal-rich ([Fe/H]$>$0, blue) and metal-poor ([Fe/H]$<$0, red) stars. Note that observed planets in (B) with eccentricities approaching the eccentricity envelope (blue band on plot) preferentially come from systems with super-solar metallicity ([Fe/H]$>$0), which is matched by our simulations (D). The contours are defined as in Figure \ref{f1}.} \label{f2}
\end{figure*}

Shifting our focus to the orbital-separation distribution, several further observational features become apparent. In Figure \ref{f2}A, there is an upper limit to the eccentricities in the inner $\sim 1$ au, which we call the eccentricity envelope (black solid line). A mass segregation in semimajor axis is seen when we separate the planets into higher-mass ($\mpl >\mmed$ $\mj$, orange) and lower-mass ($\mpl < \mmed$ $\mj$, purple), when compared to the median planet mass of $\mmed \mj$ found in the observational sample. Higher-mass planets are preferentially found with $a>0.5$ au. Furthermore, dividing the planets by metallicity in Figure \ref{f2}B reveals that the planets approaching the eccentricity envelope (blue band) are more likely to come from higher-metallicity systems, following the trend from Figure \ref{f1}. In summary, the shape of the eccentricity distribution, together with its mass- and metallicity-dependent features, are important clues to the formation of these systems.

\section{Eccentricities as a Result of a Giant-impacts Phase}
The observational features that we attempt to match with our model are: (1) the eccentricity envelope, (2) the mass-eccentricity relation, and (3) the correlation between eccentricity and stellar metallicity. We do this by making a non-standard assumption that the planets are the result of collisional growth in systems initially consisting of multiple giant planets that underwent a giant-impacts phase, at least in systems where eccentricity excitation occurs. If one is willing to make that leap, all three features can be recovered. We plot our simulation results in Figure \ref{f1}C, Figure \ref{f1}D, Figure \ref{f2}C, and Figure \ref{f2}D.

\begin{figure}[ht!] 
\includegraphics[scale=.30]{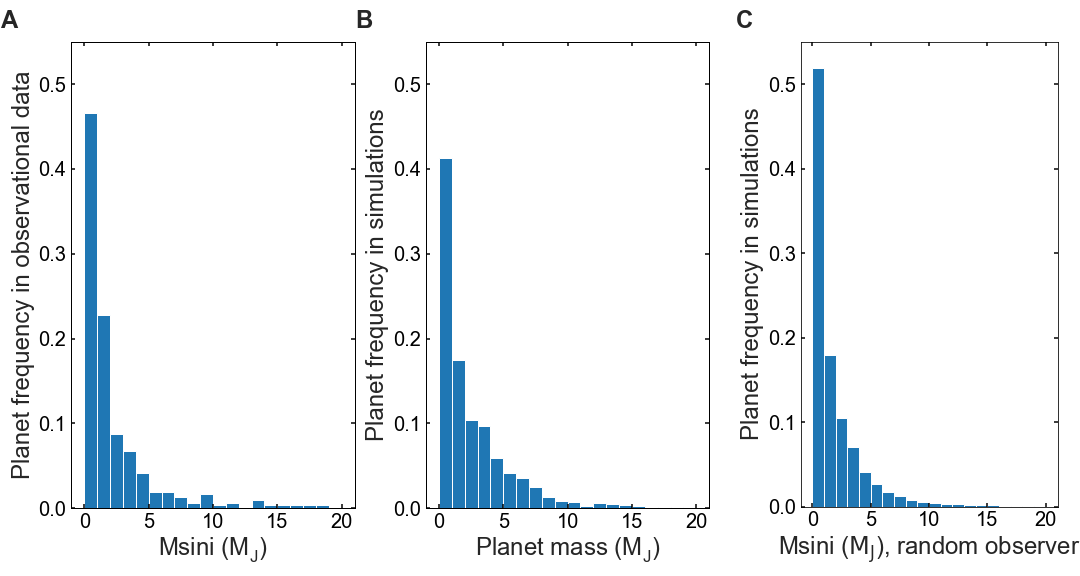}
\centering
\caption{Distribution of planet masses corresponding to Figures \ref{f1} and \ref{f2} for (A) observations and (B) \& (C) simulations. (C) displays the distribution of Msin$i$ in our simulations that would be measured by the radial-velocity method, as seen from a random ``observer" (see Section \ref{methods})} \label{f3}
\end{figure}

Our proposed giant-impacts phase is analogous to a dynamical instability that occurred in Solar System formation, where collisions assembled the inner, rocky planets. In general, a gravitational encounter between two protoplanets leads to either a physical collision or a scattering that results in a velocity deflection. Which is statistically more likely to occur depends on the relative velocities of the bodies. Recall that the collision cross section goes as $\sigma_\mathrm{coll} \sim \pi R^2 (1+\vescp^2/v^2)$, while the strong scattering cross section goes as $\sigma_\mathrm{scat} \sim \pi R^2 (\vescp/v)^4$. Here, $R$ is the physical radius, $\vescp$ is the planet's escape velocity, and $v$ is the protoplanets' relative velocity. Thus, for small velocities $v<\vescp$, $\sigma_\mathrm{coll} < \sigma_\mathrm{scat}$ and scattering dominates. As $v$ increases and approaches $\vescp$, the cross section for strong scattering approaches the physical cross section, $\pi R^2$, which is always less than or approximately equal to $\sigma_\mathrm{coll}$. Therefore, for sufficiently large $v$, $\sigma_\mathrm{coll} > \sigma_\mathrm{scat}$, and the planets are more likely to collide than be deflected further \citep[e.g.][]{petr2014}. This limits the $v$ that can arise from scattering to, at most, $\sim \vescp$. The Solar System's inner planets have $ \vescp$ ($\sim$ 11 km/s for Earth) less than the orbital escape velocity, $ v_\mathrm{esc, star}$ ($\sim$ 42 km/s at 1 au). In general, they cannot readily eject each other through gravitational encounters, allowing for the existence of a giant-impacts phase. Conversely, the Solar System's giant planets have $ \vescp$ ($\sim$ 24 km/s for Neptune, $\sim$ 60 km/s for Jupiter) larger than $ v_\mathrm{esc, star}$ ($\sim$ 8 km/s at Neptune, $\sim$ 19 km/s at Jupiter), making a giant-impacts phase unlikely. However, exoplanetary systems hosting close-in giants are in the regime where $ v_\mathrm{esc, star} >  \vescp$, allowing giant impacts to leave a detectable imprint on the giant exoplanet mass distribution (Figure \ref{f3}).

\section{Methods} \label{methods}
To generate our numerical results, we perform N-body simulations using Mercury6 \citep{chambers}, including growth by planet-planet collisions with mass and momentum conservation. Our planet radii follow nominal mass-radius relations constructed from the literature \citep{fortney}. After each collision, we calculate the planet's mass and reassign its radius using these expressions. The reason we do this is because the escape velocity is critical for the dynamical evolution. The radii for planets below 10 Earth masses ($M_\earth$) were calculated using the mass-radius relation for a rocky core in the absence of a gas envelope, \citet[Equation 7]{fortney}. Planets with masses between 10 and 200 $\mearth$ were assigned radii based on the theoretical models of \citet[Table 2]{fortney} for a 10 $\mearth$ core surrounded by a H/He envelope at 300 Myr. Planets above 200 $\mearth$ were assigned 1.1 Jupiter radii ($\rj$), which is consistent with the models, which show that the radius of Jupiter-sized planets stays roughly fixed above a Jupiter mass.
The computed radii $R$, for planets of mass $\mpl$, in $\mearth$, are as follows:

\begin{equation}
R = \begin{cases} 
(0.16(\text{log}_{10}\mpl)^2+0.74\text{log}_{10}\mpl+1.12) \re &  \mbox{if } \mpl<10 \mearth \\ 
(-1.0 \times 10^{-6} \mpl^4 + 1.8 \times 10^{-4} \mpl ^3 -1.1 \times 10^{-2} \mpl^2 +0.28 \mpl - 1.7) \rj &  \mbox{if } 10 < \mpl < 28 \mearth \\
(-7.2\times10^{-11}\mpl^4 + 8.1\times10^{-8}\mpl^3-3.2\times10^{-5}\mpl^2 + 5.5\times10^{-3}\mpl + 0.8) \rj &  \mbox{if } 28 < \mpl < 200 \mearth \\
1.1 \rj &  \mbox{if } \mpl>200 \mearth.
\end{cases}
\end{equation}

The planet densities $\rho$ in the simulations were then set to be: $\rho = 3M/(4\pi R^3)$. We note that our mass-radius relation is only an approximation. Collisions cause inflation and they may not result in a perfect merger. For example, \citet{hwang} have shown that collisions between sub-Neptune-sized bodies may result in significant atmospheric mass loss. Given these results, collision outcomes between gas giant planets merit future study.

For all of the simulations presented in this Letter, the initial eccentricities were set to 0, the mean anomalies and longitudes of the ascending node were drawn randomly from a uniform distribution between 0 and 360 degrees. We define the limits of the simulation to be at 1000 au from the host star, beyond which planets are removed from the simulation. The initial timestep was set to be the minimum of 3 days or 1/15 of the orbital period of the closest-in planet, and the accuracy parameter was set to $10^{-12}$. We set the radius of the central body to 0.005 AU, the mass to 1 Solar mass, and the Hybrid integrator changeover distance to 3 Hill radii. All simulations were run for $2 \times 10^7$ years, a reasonable time beyond which instabilities become much less frequent \citep{chatterjee,juric}. The eccentricities reach a statistical equilibrium after this timescale \citep{juric}. We found that our features of interest were already imprinted at 1 Myr, and persisted at 5, 10, and 15 Myr snapshots. We expect that integration for the several-Gyr ages, typical of observed systems, would reduce the final number of planets, but only by a modest amount.

The main results of this work are presented for a set of 694 numerical integrations, the initial inclinations $i$ for which were drawn randomly from a uniform distribution between 0 to 1 degree.

We call the initial total mass in planets in each system the ``disk mass". In constructing our simulations, we first considered a single disk mass. Each disk mass has a corresponding eccentricity envelope, which is higher for greater disk masses because they typically form higher-mass planets. Thus, the disk mass is what determines how high the eccentricities can get. As a result, higher-mass planets have higher eccentricities across systems, contrary to our expectation for a single system. The correlation of eccentricity and metallicity in the observational data suggests that the observed distribution reflects a collection of varied disk masses. Therefore, to reproduce the observational data, we need to construct a set of simulations with initial disk masses drawn from a distribution. We do not know a priori what this distribution is. However, if we choose a distribution that is a few times the observed planet mass distribution, our (Figure \ref{f3}) set of systems evolves to match the observed distribution of planet masses, after ejections and mergers. We choose to draw the disk masses $\mdisk$ from the exponential distribution,

\begin{equation} \label{dist}
f(\mdisk/\mj,1/\beta)=1/\beta \mathrm{e}^{-\mdisk/ (\beta \mj)},
\end{equation}
with the scale parameter $\beta=20$ set to produce the planet mass distribution. We re-draw any disks with masses $<0.1 \mj$, as our disks need to be able to produce giant planets. An upper disk mass cutoff of 50 $\mj$ is employed. Our simulations are not otherwise tuned. Each disk mass is allocated into planets in the following way: 10 planet masses are first drawn from a uniform distribution from 0 to 1, then each planet mass is scaled, such that the sum of the planet masses in each system is equal to the disk mass. The planets are uniformly distributed in log($a$) between 0.03 and 10 au. To justify the logarithmic spacing in semimajor axis, we recall that the number of planetary Hill radii, $\rh$, that can dynamically fit into a given semimajor axis range is: $a/\rh = (\mpl/(3M_\mathrm{star}))^{-1/3}$. Here, the Hill radius, $\rh = a(\mpl/(3M_\mathrm{star}))^{1/3}$, sets the distance scale from the planet at which gravitational interactions with other planets become strong. Since this ratio is independent of $a$, each successive bin in log($a$) from the star can fit the same number of planetary $\rh$ as the one prior, and our uniform spacing in log($a$) is appropriate.

As long as the planets were dynamically spaced close enough to interact, the final outcome was not significantly dependent on the way we distributed the disk mass into planets. For example, a 20 $\mj$ disk, allocated into 7 planets, each with mass 2.9 $\mj$, produced similar results to a 20 $\mj$ disk allocated into 50 planets with mass 0.4 $\mj$. Both typically produce planets with masses up to $\sim$ 10 $\mj$ after collisions. We chose 10 planets to allow the planets to have multiple close encounters within the first $\sim$ 10 thousand years of the simulation, after which the number of planets decreased. We have found, in agreement with previous work \citep[e.g.][Figures 1 and 11]{juric}, that although the planet multiplicity generally dropped further as we extended the time to $10^8$  years, and the individual planet orbital properties varied with time, our overall features of interest in the distributions of eccentricity with mass and semimajor axis persisted. The final planet mass distribution from the simulations (Figure \ref{f3}B) was consistent with the mass distribution in the observational sample (Figure \ref{f3}A), suggesting that our starting guess for the disk masses was reasonable. We note that planet migration before the gas disk dissipates may mean that fewer planets per system may be required, if it causes planets to migrate in and experience dynamical instability.

 For our simulation results, plotted in Figures \ref{f1}C, \ref{f1}D, \ref{f2}C, \ref{f2}D, \ref{f3}B, and \ref{f3}C, we consider planets observable if they have $a<6$ au and radial velocities $>$2 m/s, which roughly matches the typical precision of radial-velocity surveys contributing to our observational sample \citep{butler2006}. Radial velocities, $K$, are estimated by $28.4
 (M_\mathrm{p} /\mathrm{M}_\mathrm{J})(T/\mathrm{yr})^{1/3}$  m/s, with $M_\mathrm{p}$ - the planet mass and $T$ - the orbital period \citep{lovis}. We comment that including the eccentricity dependence $(1-e^2)^{-1/2}$ in the expression for $K$ does not significantly change our results. To determine the likelihood of observing systems with multiple planets, we ``observe" each resulting system of our set of simulations from 1000 random directions. We evaluate the observability of each planet by computing the radial velocity component along the line of sight to our random ``observer", i.e. we multiply the mass of each planet by $\mathrm{sin} i$ in the radial-velocity formula. We find that $\sim 70 \%$ of systems have an observable multiplicity of 1 or 2, $\sim 20 \%$ have no observable planets per system, and $\sim 10 \%$ have 3 or more. On average, $\sim$ 1.6 planets can be observed per system. Our simulated distribution of Msin$i$, generated by observing each system from random angles as described above, is shown in Figure \ref{f3}C.

Trends with stellar metallicity that arose in the observed populations were of particular interest (Figures \ref{f1} and \ref{f2}). Stellar metallicity is not a parameter in our simulations. However, we can make a simple assumption that a higher total planet mass is correlated with stellar metallicity, which is reasonable given that high-metallicity stars are more likely to host close-in planets \citep{fischer}, and eccentric warm Jupiters are preferentially found in high-metallicity systems \citep{dawson2013}. To map the synthetic distribution of disk masses containing observable planets to the observed distribution of stellar metallicities, we first assume a linear relationship between $\mathrm{log}_{10}(\mdisk/\msun)$ and $[\mathrm{Fe}/\mathrm{H}]$ of the form $[\mathrm{Fe}/\mathrm{H}] = a \, \mathrm{log}_{10}(\mdisk/\msun)+b$, where $\msun$ is the Solar mass. We then find that $a=0.4$ and $b=0.9$ in the formula give us the best fit, such that the range of observed metallicities matches the range of assigned metallicities from our simulations. Therefore, we convert the disk mass from our simulations to metallicity using the following relation: $[\mathrm{Fe}/\mathrm{H}] = 0.4 \, \mathrm{log}_{10}(\mdisk/\msun)+0.9$. We note that we do not expect there to exist a perfect one-to-one correspondence between stellar metallicity and disk mass, and our nominal relation is only suggestive.

\section{Discussion} \label{disk}
Figure \ref{f1}B displays the distribution of eccentricities as a function of planet mass in our simulations. Contours correspond to enclosed probabilities. The upper contour encloses 90\% of the points, and each successive contour below is a 10\% decrement in enclosed probability from the one above. The contours are defined the same throughout all of the figures in this Letter, to allow one-to-one comparison of the simulations and the observations. We remind the reader that the only tuning required was the choice of an initial disk mass distribution to match the observed planet mass distribution, and a sufficient number of planets to ensure dynamical interaction. Our simulations in Figure \ref{f1}D reproduce the observational feature seen in Figure \ref{f1}C: $e > 0.6$ planets have $\mpl > 0.5$ $\mj$ and [Fe/H]$>$0. Higher-metallicity systems have more mass available for planet formation, which leads to more and larger planets that can excite each other to higher eccentricities.

In Figure \ref{f2}, both the observations (A) and the simulations (C) are bounded by the eccentricity envelope. This is the upper limit to the eccentricities produced by planet-planet scattering, which deflects the planetary velocities in random directions. The orbit deviation from circular is often called the “random” velocity, and can be approximated as $v_{\mathrm{r}}=ev_{\mathrm{K}}$, where $v_{\mathrm{K}}=(\mathrm{G}M_{\mathrm{star}}/a)^{1/2}$ is the orbital Keplerian velocity, G - the gravitational constant, and  $M_{\mathrm{star}}$ - the stellar mass \citep{gls}. Since scattering can only excite $v_{\mathrm{r}}$ up to $\vescp$ (Section 4), the eccentricity is limited to: $v_{\mathrm{r}}=ev_{\mathrm{K}}\approx e\vescp$. This yields a maximum eccentricity of $e=\vescp/v_{\mathrm{K}}
\approx \vescp/v_{\mathrm{esc,star}}$ . Here, $v_{\mathrm{esc, star}}=(2\mathrm{G}M_{\mathrm{star}}/a)^{1/2}$  is the escape velocity from the star at the planet's $a$, which is comparable to $v_{\mathrm{K}}$ . The eccentricity envelope curves in Figure \ref{f2} are
\begin{equation} \label{envelope}
e = \vescp/v_\mathrm{{esc, star}} = (\mpl a/(R M_\mathrm{star}))^{1/2} ,
\end{equation}
shown for $\mpl = \mmed \mj$, the median observed planet mass. Planets in higher-metallicity systems are scattered to a greater range of eccentricities than those with lower metallicity (blue band in Figure \ref{f2}B and D). 

\begin{figure*}[ht!]
\plotone{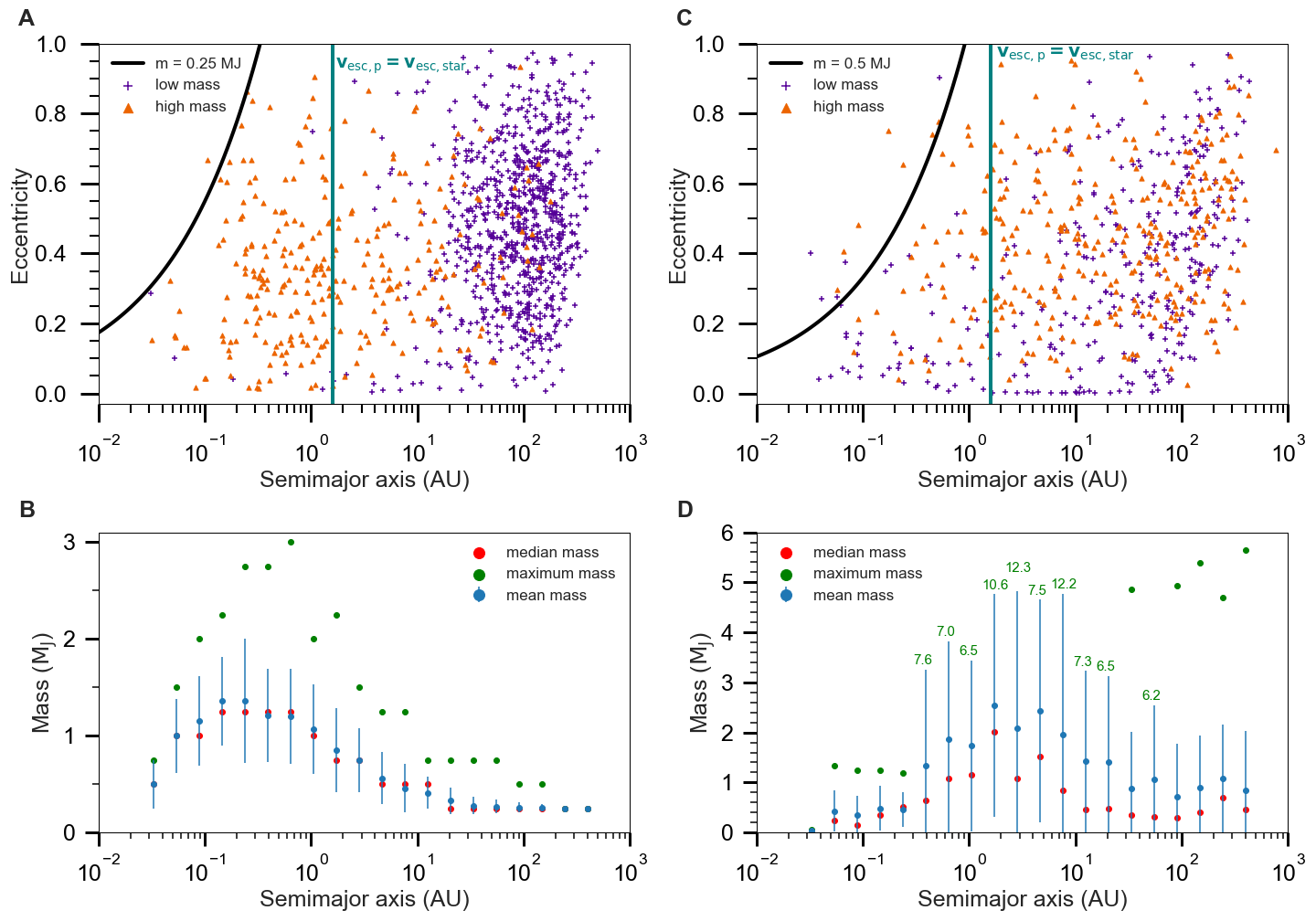} 
\caption{Eccentricities and masses vs. semimajor axis for simulations starting with 20 planets in the semimajor axis range 0.03 to 100 au. (A) Initial planet masses are all 0.25 $\mj$. The inner, higher-mass planets (orange triangles) grew via planet-planet collisions. The lower-mass (purple pluses), outermost planets never collided. The vertical teal line at 1.6 au marks where the escape velocity of a 0.25 $\mj$ Saturn-radius planet is equal to the orbital escape velocity of the planet from the star. (B) shows the corresponding distribution of planet masses, including the median (red dot), mean (blue dot), and maximum (green dot) masses. The vertical blue bars mark one standard deviation about the mean. Most high-mass planets form interior to the distance where $\vescp=v_\mathrm{esc,star}.$ (C) and (D) provide corresponding plots to (A) and (B), respectively, for a set of initial disk masses designed to match the distribution used in Figures \ref{f1}-\ref{f3} (see text), but adjusted to incorporate the additional bin in $\mathrm{log}_{10} (a)$ compared to the results shown in Section 5. In (D), the values for the maximum planet masses that extend beyond the axis range are printed in green by the error bars corresponding to each bin in semimajor axis. The highest-mass products of collisional growth have $1 < a < 8$ au.} \label{f4}
\end{figure*}

Because we are interested in the implications of planet-planet scattering for planets at all semimajor axes, not just those that are currently observable, we perform two additional sets of simulations with 20 planets uniformly distributed in log($a$) between 0.03 and 100 au. The outcome for a single disk mass of 5 $\mj$ is presented in Figures \ref{f4}A and B for a set of 192 simulations. For each realization of the 20-planet systems, we re-drew the semimajor axes uniformly in log($a$), the inclinations uniformly from 0 to 1 degree, and the mean anomalies and longitudes of the ascending node uniformly from 0 to 360 degrees. In Figure \ref{f4}A, the higher-mass products of planet-planet collisions are seen to the left of the teal solid line, which marks $\vescp=v_\mathrm{esc, star}$. Lower-mass planets are seen to the right, where scattering is a more likely outcome of planet-planet interactions. The detailed mass distribution is shown in (B). Green, red, and blue dots show the maximum, median, and mean masses, respectively. The vertical blue lines display one standard deviation about the mean. The highest-mass planets typically form interior to the distance where $\vescp = v_\mathrm{{esc, star}}$.

The result for a distribution of disk masses is presented in Figure \ref{f4}C and D for a set of 175 simulations. We draw the disk masses from the exponential distribution (Equation \ref{dist}), scaled by a factor of 1.4 to account for the additional bin in log($a$). We employed an upper disk mass cutoff of 70 $\mj$. The 1-8 au region hosts planets of higher mass. Interestingly, adding mass to the outer disk reduces the number of planets in the inner regions of higher-mass disks. The mass needed beyond 10 au to match the mass segregation feature from Figure \ref{f2}A, while reproducing the metallicity feature from Figure \ref{f2}B, merits future study.

\begin{figure*}[ht!]
\plotone{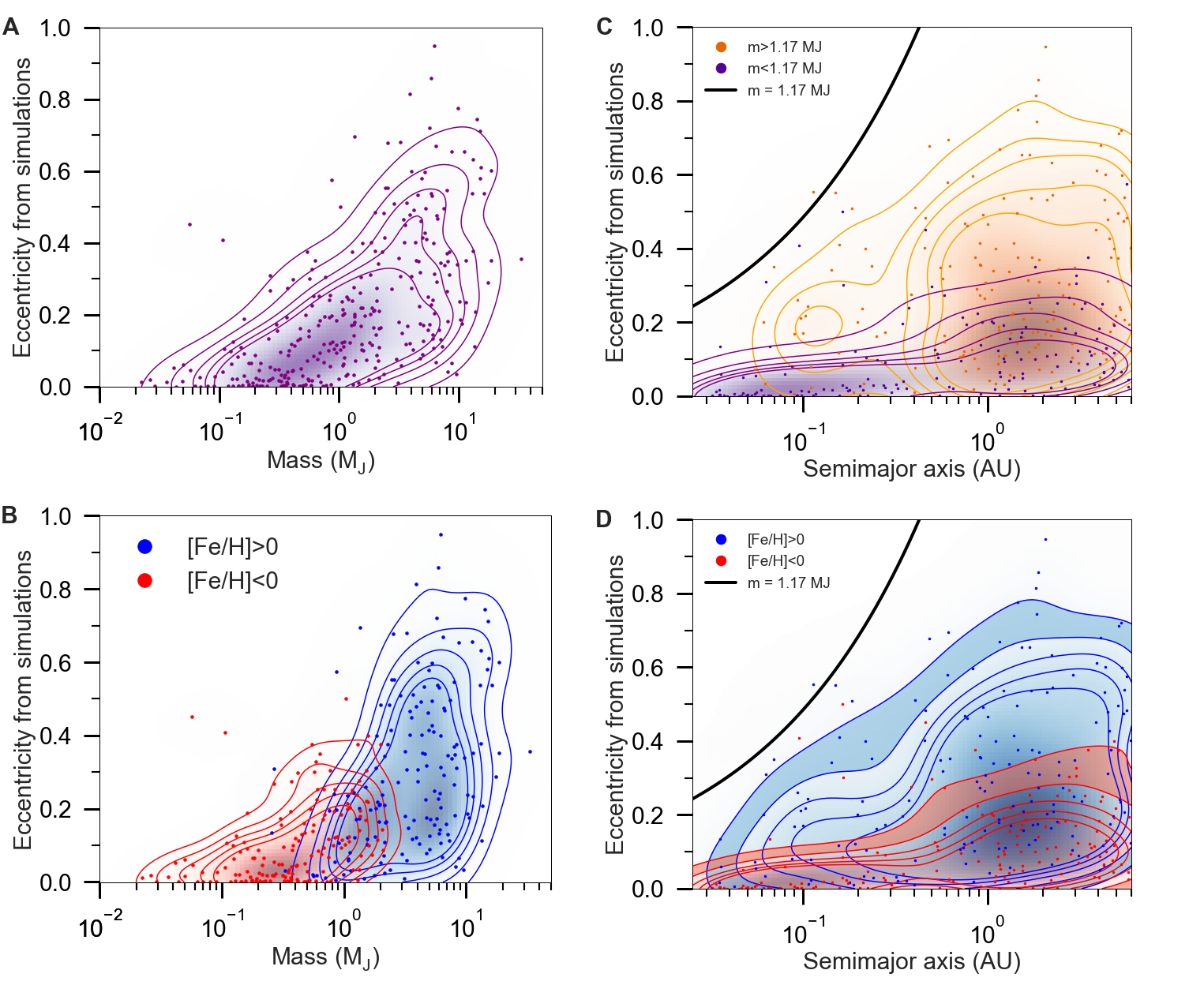} 
\caption{(A) Our 2D simulations reproduce the correlation between mass and eccentricity in the distribution. (B) Planets orbiting high-metallicity ([Fe/H]$>$0, blue) stars are able to be excited to a higher eccentricities than planets orbiting low-metallicity stars ([Fe/H]$<$0, red). Metallicities are used as a proxy for the initial disk masses (Section \ref{disk}). (C) and (D) Distribution of eccentricities as a function of semimajor axis in our 2D simulations. In (C), the planets of higher mass (orange) occur more frequently beyond $\sim$0.5 au, and can reach higher eccentricities than the lower-mass (purple) planets, which are confined to $e<0.4$. In (D), planets orbiting high-metallicity stars are excited to higher eccentricities than planets orbiting low-metallicity stars throughout the entire semimajor axis range where they occur. The contours are defined as in Figure \ref{f1}.} \label{f5}
\end{figure*}

We find it illustrative to repeat our main simulations in two dimensions, to avoid potential complications from Kozai oscillations. We start out with 10 planets using the same set of initial disk masses as described in Section \ref{methods}, except we initialize the inclinations to 0. The starting semimajor axes of the planets are randomly distributed in log($a$) from 0.03-1 au, and from 1-10 au, with the density reduced by a factor of 5 in the inner region. This selection is appropriate to match the observed mass-segregation feature in Figure \ref{f2}A. We plot the results for this set of 180 simulations in Figure \ref{f5}, for direct comparison to Figures \ref{f1} and \ref{f2}. The features of interest are recovered in these two-dimensional simulations, suggesting that they are a result of planet-planet scattering. In the absence of mutually inclined orbits, the planets experience more collisional growth, especially in higher-mass disks, where the planets are dynamically spaced close enough to be more likely to interact. The smaller planets in the higher-mass disks get absorbed by collisions more frequently (Figure \ref{f5}A) than the mutually inclined planets in Figure \ref{f1}C. The eccentricity vs. semimajor axis distribution in Figure \ref{f5}D reproduces the observational feature seen in Figure \ref{f2}B: the planets from higher-metallicity systems extend above the lower-metallicity region in a blue band on the plot. We plot the same distribution, instead dividing the planets by mass, in Figure \ref{f5}C. The initial reduction in planet number density in the inner region allowed us to reproduce the mass segregation feature seen in the observations (Figure \ref{f2}A). This mass reduction may be appropriate if giant planets do not form as easily in the inner 1 au of protoplanetary disks. On average, $\sim$1.8 observable planets are left per system in our 2D simulations.

\section{Conclusion}
We have suggested a giant-impacts phase in the evolution of giant exoplanetary systems, which creates a population of higher-mass planets in the collisional growth region (peaking at $\sim 3$ au for a Solar-mass star; see Figure \ref{f4}D). Beyond, we predict a population of lower-mass planets that avoided mergers, some of which were scattered out on high-eccentricity orbits. This is consistent with the results of the Gemini Planet Imager Exoplanet Survey (GPIES) \citep{nielsen}, who found that planets more massive than $\sim 3 \mj$ around Solar-type stars are mostly found in the radial-velocity regime ($a <$ 5 au), rather than in the direct-imaging regime ($a >$10 au). Future microlensing and direct imaging surveys, sensitive to finding planets outside of the distance limits of radial-velocity surveys, will enable further observational tests of this predicted mass separation in the outer giant exoplanet population.

As evidence of mergers could correlate with orbital eccentricities, we speculate that inflated Jupiters would exist on preferentially eccentric orbits. Though collisions can happen on any timescale, most occur early, so this correlation would be most likely observable for stars with ages less than the $\sim$ 100 Myr cooling time for giants \citep{fortney}. In addition, since some warm Jupiters have substantial cores \citep{thorn}, it is reasonable to suggest that early-on in their evolution planetary systems could have consisted of multiple planets, which experienced collisional growth.

We thank Jonathan Fortney, Yanqin Wu, and Kassandra Anderson for helpful discussions. We thank Bruce Macintosh for pointing us to the GPIES results of \citet{nielsen}. This work made use of the Exoplanet Orbit Database at exoplanets.org. RMC and RF are supported by NSF CAREER grant AST-1555385.

\end{document}